\documentclass[amsmath,amssymb,prc,final,showpacs,twocolumn]{revtex4}

\usepackage{bm,hyphenat,xspace}
\usepackage{graphicx,epsfig}
\usepackage{color}

\newcommand {\mbf}[1]{{\mathbf{#1}}}

\newcommand {\mcu}{\mathcal{U}}

\newcommand{\He}{{}^3\mathrm{He}}
\newcommand{\Heq}{{}^4\mathrm{He}}

\newcommand{\nH}{n\text{-}{}^3\mathrm{H}}
\newcommand{\pHe}{p\text{-}{}^3\mathrm{He}}
\newcommand{\pH}{p\text{-}{}^3\mathrm{H}}
\newcommand{\nHe}{n\text{-}{}^3\mathrm{He}}

\newcommand{\dd}{d\text{-}d}

\newcommand{\bmx}{{\bm x}}
\newcommand{\bmy}{{\bm y}}
\newcommand{\bmz}{{\bm z}}
\newcommand{\nuc}[2]{\ensuremath{\rm{^{#1}}#2}}  
\definecolor {red}{rgb}{1.00,0.00,0.00}

\begin{document}

\title {Benchmark calculation of $\nH$ and $\pHe$ scattering}

\author{M.\ Viviani$^{\,{\rm a}}$, A. Deltuva$^{\,{\rm b}}$,
  R. Lazauskas$^{\,{\rm c}}$,  J. Carbonell$^{\,{\rm d}}$,
  A.~C.~Fonseca$^{\,{\rm b}}$, A.\ Kievsky$^{\,{\rm a}}$,
  L.E.\ Marcucci$^{\,{\rm  e,a}}$, and S. Rosati$^{\,{\rm  e,a}}$}

\affiliation{
$^{\,{\rm a}}$ INFN-Pisa, 56127 Pisa, Italy\\
$^{\,{\rm b}}$ Centro de F\'{\i}sica Nuclear da Universidade de Lisboa,   P-1649-003 Lisboa, Portugal \\
$^{\,{\rm  c}}$ IPHC, IN2P3-CNRS/Universit\'e Louis Pasteur BP 28, F-67037 Strasbourg Cedex 2, France\\
$^{\,{\rm  d}}$ CEA-Saclay, IRFU/SPhN, F-91191 Gif-sur-Yvette, France\\
$^{\,{\rm  e}}$ Department of Physics, University of Pisa, 56127 Pisa, Italy
}

\received{\today}

\begin{abstract}
The  $\nH$ and $\pHe$ elastic phase-shifts  below
the trinucleon disintegration thresholds are calculated by solving the
4-nucleon problem with three different realistic nucleon-nucleon interactions (the
I-N3LO model by Entem and Machleidt, the Argonne $v_{18}$ potential
model, and a low-$k$ model derived from the CD-Bonn potential). Three
different methods -- Alt, Grassberger and Sandhas, Hyperspherical
Harmonics, and Faddeev-Yakubovsky -- have been used and their
respective results are compared. For both $\nH$ and $\pHe$ we observe
a rather good agreement between the three different theoretical
methods. We also compare the theoretical predictions with the
available experimental data, confirming the large underprediction of
the $\pHe$ analyzing power.
\end{abstract}

 \maketitle


\section{Introduction}
\label{sec:intro}

The four--nucleon (4N) system has been object of intense studies
in recent years. In first place, this system is particularly interesting as a
``theoretical laboratory" to test the accuracy of our present
knowledge of the nucleon--nucleon (NN) and three nucleon (3N) interactions.
In particular, the effects of the NN P-waves and of the 3N force  are
believed to be larger than in the $A=2$ or $3$ systems. Moreover,
it is the simplest system where the 3N interaction in  channels
of total isospin $T=3/2$ can be studied. In second place, there is a number of reactions
involving four nucleons which are of extreme importance for astrophysics, energy
production, and studies of fundamental symmetries. As an example, reactions
like $d + d \rightarrow \nuc{4}{He} + \gamma$ or $p + \nuc{3}{He} \rightarrow
\nuc{4}{He} + \nu_{e} + e^{+}$ (the $hep$ process) play important roles in
solar models and in the theory of big-bang nucleosynthesis.

Nowadays, the 4N bound-state problem can be numerically solved with
good accuracy. For example, in Ref.~\cite{Kea01} the binding energies and
other properties of the $\alpha$-particle were studied using the
AV8$'$~\cite{AV18+} NN interaction; several different techniques produced
results in very close agreement with each other (at the level of less than
1\%). More recently, the same agreement has also been obtained  considering
different realistic NN+3N force models~\cite{Wea00,Nogga03,Lazaus04,Viv05}.

In recent years, there has also been a rapid advance in solving the 4N
scattering problem with realistic Hamiltonians. Accurate calculations of four-body
scattering observables have been achieved in the framework of the
Alt-Grassberger-Sandhas (AGS) equations~\cite{DF07,DF07b,DF07c,DF10,DFS08}, solved in momentum
space, where the long-range Coulomb interaction is treated using the
screening and renormalization method \cite{Alt78,DFS05}.
Also solutions of the Faddeev-Yakubovsky (FY) equations in
configuration space~\cite{Cie98,Rimas_03,LC_FB17_03,Lea05,Lazaus09} and the application of the
Hyperspherical Harmonics (HH) expansion method~\cite{rep08} to the solution
of this problem have been reported~\cite{Vea09,Vea10}.

In addition to these methods, the solution of the 4N scattering
problem has been obtained also by using the resonating group model
(RGM) method~\cite{HH97,PHH01,HH03,Sofia08}.
Calculations of scattering observables using the Green's function
Monte Carlo method are also underway~\cite{Wiringapc}.

The 4N studies performed so far have evidenced several discrepancies between
theoretical predictions and experimental data. Let us consider
first $\nH$ elastic scattering. Calculations based on NN
interaction models disagree~\cite{Lazaus04,DF07,Vea09}
rather sizeably with the measured total cross section~\cite{PBS80}, both at
zero energy and in the ``peak'' region ($E_n\approx 3.5$ MeV). This
observable is found to be very sensitive to the NN interaction
model~\cite{DF07}. At low energy, the discrepancy is removed by including a 3N
force fixed to reproduce the triton binding energy~\cite{VKR98,Cie98,PHH01,Vea09}, but it
remains in the peak region. The analysis of
the differential cross section has shown similar discrepancies, but definitive conclusions are
difficult to extract since the experimental errors are rather large.

In this respect, the $\pHe$ elastic scattering is more interesting  since
there exist several accurate measurements of both the unpolarized cross
section~\cite{Fam54,Mcdon64,Fisher06} and the proton analyzing power
$A_{y0}$~\cite{All93,Vea01,Fisher06}. The calculations performed so far (with a
variety of NN and NN+3N interactions) have shown a large discrepancy
between theory and experiment for $A_{y0}$~\cite{Lea05,Fisher06,DF07b,Vea01,Fon99}.
In addition, at the Triangle Universities Nuclear Laboratory (TUNL),
there has been recently a new set of accurate
measurements of other $\pHe$ observables (the $\He$ analyzing power
$A_{0y}$ and some spin correlation observables as $A_{yy}$, $A_{xx}$,
$A_{xz}$, $A_{zx}$, and $A_{zz}$)
at $E_p=1.60$, $2.25$, $4.05$ and $5.54$ MeV,
which has allowed a phase-shift analysis (PSA)~\cite{Dan10}.
A preliminary comparison with these data has been reported in
Ref.~\cite{Vea10}.

In order to have definite answer about the ability of the different interaction
models to reproduce the experimental data it is certainly of interest to
establish the accuracy reached by the
theoretical methods in the solution of the $A=4$ scattering problem.
In a previous benchmark, the results obtained by different groups
working with different techniques  were found to be
at variance with each other~\cite{Lea05}.
Clearly, this situation should be clarified before questioning the
ability of present NN+3N force models to describe the experimental
data beyond the binding energy of $^4$He. This is the purpose of the
present paper, in which we present low energy $\nH$ and $\pHe$ scattering results
obtained by three different groups, using independent methods to
solve the four-body problem, i.e., the AGS
equations, the variational HH expansion, and the FY
equations.

The potentials used in this paper are the I-N3LO model by Entem and
Machleidt~\cite{EM03}, with cutoff $\Lambda=500$ MeV, the Argonne $v_{18}$
(AV18) potential model~\cite{AV18}, and a low-$k$ model derived from the
CD-Bonn potential~\cite{BKS03}. The I-N3LO
potential has been derived using an effective field theory approach
and the chiral perturbation theory up to
next-to-next-to-next-to-leading order.
The AV18 potential is a phenomenological potential having
a rather strong repulsion at short interparticle
distances. The low$-k$ potentials have been obtained
separating the Hilbert space into low and  high momentum regions and
using the renormalization group method~\cite{BKS03} to
integrate out the  high-momentum components
above a cutoff $\Lambda$. The low$-k$ potential adopted in this work
is obtained starting from the realistic CD-Bonn potential~\cite{Mach01}
and using a smooth cutoff $\Lambda=2.5$ fm$^{-1}$. The cut of the
high-momentum part is reflected in configuration space in an almost total
absence of the repulsion at short interparticle distances. Note that
the first and third model are non-local, while AV18 is local in
configuration space. The three potentials reproduce equally well the $np$ and
$pp$ data, and are a representative set of the large variety of
modern NN potential models. We note finally that I-N3LO and
AV18 interactions, without the inclusion of a suitable 3N interaction
model, largely underestimate the $\Heq$ binding energy $B(\Heq)$. On the
contrary,  with the adopted low-$k$ potential model we have
$B(\Heq)=29.04$ MeV, slightly
overestimating the experimental value of $28.30$ MeV.

This paper is organized as follows. In Section~\ref{sec:methods}, a brief
description of the methods used for this benchmark is reported. In
Section~\ref{sec:results}, a comparison  between the results obtained
within the different schemes is shown. In Section~\ref{sec:comp-theo-expt}, the
theoretical calculations are compared with the available experimental data.
The conclusions will be given in Section~\ref{sec:conc}.

\section{Methods}
\label{sec:methods}
In order to solve the 4N scattering problem we employ the
AGS equations, the HH method, and the FY equations.
The various procedures are briefly described below.

The total kinetic energy, $T_{c.m.}$,
in the center of mass (c.m.) and the nucleon kinetic energy,
$E_N$ ($N=p$, $n$), in the laboratory reference frame are given by
\begin{equation}
  T_{c.m.}={q^2\over 2\mu}\ , \qquad
  E_N={4\over 3} T_{c.m.}\ ,\label{eq:energy}
\end{equation}
where $\mu=(3/4)M_N$ is the reduced mass of the $1+3$ system, $M_N$ is the
nucleon mass, and $q$ the magnitude of the relative momentum between
the two clusters.

\subsection{AGS Equations}
\label{sec:AGS}

The AGS equations \cite{grassberger:67} are integral equations
for the four-body transition operators. They are well-defined only with
short-range potentials. Nevertheless, together with the
screening and renormalization method  \cite{deltuva:05a,DF07b},
they can be applied also to the reactions involving charged particles.
In the 4N system we use the isospin formalism and solve the
symmetrized form of  the AGS equations \cite{DF07}. In this case there are only two
distinct four-particle partitions, one of the $3+1$ type and
one of the $2+2$ type. We choose those partitions to be (12,3)4 and
(12)(34) and denote them in the following by $\alpha =1$ and $2$,
respectively. The corresponding transition operators $\mcu_{\beta\alpha}$
for the initial states of the $3+1$ type, as appropriate
for the $\nH$ and $\pHe$ scattering, obey the integral equations
\begin{eqnarray}
\mathcal{U}_{11}  &= {}& -(G_0 \, T  G_0)^{-1}  P_{34}
 - P_{34} \, U_1\, G_0 \, T G_0 \, \mathcal{U}_{11}\nonumber\\
   && \qquad  + {U}_2   G_0 \, T G_0 \, \mathcal{U}_{21} ,
\label{eq:U11} \\
\mathcal{U}_{21}  &= {}&  (G_0 \, T  G_0)^{-1} \, (1 - P_{34})\nonumber\\
 &&\qquad + (1 - P_{34}) U_1\, G_0 \, T  G_0 \,
\mathcal{U}_{11} .  \label{eq:U21}
\end{eqnarray}
Here $G_0 = (E+i\epsilon -H_0)^{-1}$ is the free resolvent, $E$ being the energy of the 4N system
and $H_0$ the free Hamiltonian, and
$P_{ij}$ is the permutation operator of particles $i$ and $j$.
The (short-range) two-nucleon potential $V^s$ enters the AGS equations via the
two-nucleon transition matrix $T=V^s+V^s G_0T$
and the 3+1 and 2+2 subsystem transition operators
\begin{equation}
\label{eq:U}
U_{\alpha}  = {}  P_\alpha G_0^{-1} + P_\alpha \, T G_0 \, U_{\alpha}, \\
\end{equation}
where $P_1  = {}  P_{12}\, P_{23} + P_{13}\, P_{23}$ and
$P_2  = {}  P_{13}\, P_{24}$.
The 3+1 elastic scattering amplitudes are given by
$ \langle \mathbf{p}_f| \mathcal{T} |\mathbf{p}_i \rangle
= 3\langle \Psi_1(\mathbf{p}_f) | \mathcal{U}_{11} | \Psi_1 (\mathbf{p}_i)\rangle $
where the factor 3 results from the symmetrization and
$| \Psi_\alpha (\mathbf{p}_j) \rangle $  are properly normalized
 initial/final channel state Faddeev components.

In order to include the Coulomb interaction $V^C$
between the protons
in the $\pHe$ scattering we
use the screening and renormalization approach \cite{deltuva:05a,DF07b}.
We add to the nuclear $pp$ potential the screened Coulomb one
$V^R(r)= V^C(r) \, \exp{(-(r/R)^n)}$.
Thus, the AGS equations with $V^s+V^R$ are well-defined
but all  transition operators and the resulting amplitudes
depend on the screening radius $R$.
The renormalization procedure in the $R \to \infty$ limit yields the full $\pHe$
transition amplitude
\begin{gather}      \label{eq:T}
  \begin{split}
    \langle \mathbf{p}_f| \mathcal{T}_{(C)} |\mathbf{p}_i \rangle  = {} &
    \langle \mathbf{p}_f| t_{C}^{c.m.} |\mathbf{p}_i \rangle \\
& + \lim_{R \to \infty}
    \langle \mathbf{p}_f| [ \mathcal{T}_{(R)} - t_{R}^{c.m.} ] |\mathbf{p}_i \rangle Z_R^{-1},
    \end{split}
\end{gather}
where $\langle \mbf{p}_f| t_{C}^{c.m.} |\mbf{p}_i \rangle$ and
$ \langle \mbf{p}_f| t_{R}^{c.m.} |\mbf{p}_i \rangle$
are the proper  and screened Coulomb amplitudes between the c.m. of two charged clusters,
respectively; the former is known analytically.
The renormalization factor $Z_R$ is defined in Ref.~\cite{DF07b}. Thus,
 the long- and Coulomb-distorted short-range parts
in the scattering amplitudes are  isolated
and their infinite $R$ limit is calculated separately.
The long-range part of the amplitude $ \langle \mbf{p}_f| t_{R}^{c.m.} |\mbf{p}_i \rangle$
 is of two-body nature and its $R \to \infty$ limit after renormalization
is $\langle \mbf{p}_f| t_{C}^{c.m.} |\mbf{p}_i \rangle$.
The Coulomb-distorted short-range part $[ \mathcal{T}_{(R)} - t_{R}^{c.m.} ]$
is calculated by solving the AGS equations for $V^s+V^R$ numerically at a finite $R$ that is
sufficiently large to get $R$-independent results after the renormalization. In other words,
the $R \to \infty$ limit is reached with sufficient accuracy at finite $R$.
However, $R$ must be considerably  larger than the range of the nuclear interaction
thereby leading to a slower partial-wave convergence.
The right choice of the screening, i.e., the exponent $n$,
 is essential in dealing with this difficulty.
For a fast convergence with $R$ we have to ensure that  $V^{R}(r)$ approximates well
$V^{C}(r)$ for  $r < R$ and simultaneously vanishes smoothly but rapidly for $r>R$,
providing a comparatively fast convergence of the partial-wave expansion.
Using the optimal value $n=4$
we obtain reasonably converged results with $R$ ranging from 10 to 15 fm
and including two-proton partial waves with orbital angular momentum up to 10.
The $R$-convergence is slower at lower energies; the worst cases are the $S$ waves
at $E_p = 2.25$ MeV where we estimate the accuracy of our phase-shift results
to be around 1\%.
In contrast, the $\nH$ results are converged very well, considerably better than 0.2\%,
as demonstrated in Ref.~\cite{DF07} where also the details on the included partial waves
can be found.

\subsection{HH Expansion}
\label{sec:HH}

The wave function describing a $\nH$ or $\pHe$ scattering state with
total angular momentum quantum numbers $J,J_z$, incoming relative
orbital angular momentum $L$, and channel spin $S$
($S=0, 1$) can be written as
\begin{equation}
    \Psi_{1+3}^{LS,JJ_z}=\Psi_C^{LS,JJ_z}+\Psi_A^{LS,JJ_z} \ ,
    \label{eq:psica}
\end{equation}
where the part $\Psi_C^{LS,JJ_z}$ describes the system in the region where
the particles are close to each other and their mutual interactions
are strong. Hence, $\Psi_C^{LS,JJ_z}$ vanishes in the limit of large
inter-cluster distances. This part of the wave function is written as a linear expansion
$\sum_\mu c^{LSJ}_\mu {\cal Y}_\mu$,  where ${\cal Y}_\mu$ is a set of
basis functions constructed in terms of the HH functions
(for more details, see, for example, Ref.~\cite{rep08}).

The other part $\Psi_A^{LS,JJ_z}$ describes the relative
motion of the two clusters in
the asymptotic regions, where the $1+3$ interaction is
negligible (except eventually for the long-range Coulomb interaction). In the
asymptotic region the wave functions $\Psi_{1+3}^{LS,JJ_z}$ reduces to
$\Psi_{A}^{LS,JJ_z}$, which must therefore be the appropriate asymptotic
solution of the Schr\"odinger equation. Let us consider, for example,
the $\pHe$ case. Then, $\Psi_{A}^{LS,JJ_z}$ can be decomposed
as a linear combination of the following functions
\begin{eqnarray}
  \Omega_{LS,JJ_z}^\pm &=&  \sum_{l=1}^4
  \Bigl [ Y_{L}(\hat{\bm y}_l) \otimes  [ \phi_3(ijk) \otimes s_l]_{S}
   \Bigr ]_{JJ_z} \nonumber\\
  &&\times \left ( f_L(y_l) {\frac{G_{L}(\eta,qy_l)}{q y_l}}
          \pm {\rm i} {\frac{F_L(\eta,qy_l)}{q y_l}} \right ) \ ,
  \label{eq:psiom}
\end{eqnarray}
where ${\bm y}_l$ is the distance between the proton (particle $l$) and \nuc{3}{He}
(particles $ijk$), $q$ is the magnitude of the relative momentum between the
two clusters, $s_l$ the spin state of particle $l$, and $\phi_3$ is the $\He$
wave function.
Moreover, $F_L$ and $G_L$ are the regular and irregular Coulomb function,
respectively, with $\eta=2\mu e^2/q$.
The function $f_L(y)=[1-\exp(-\beta y)]^{2 L+1}$ in Eq.~(\ref{eq:psiom})
has been introduced to regularize  $G_L$  at small $y$, and
$f_L(y) \rightarrow 1$ as $y$ is large, thus  not affecting the asymptotic
behavior of $\Psi_{1+3}^{LS,JJ_z}$. Note that for large values of $qy_l$,
\begin{eqnarray}
  \lefteqn{ f_L(y_l) G_{L}(\eta,qy_l)\pm {\rm i} F_L(\eta,qy_l) \rightarrow
   \qquad\qquad} &&  \nonumber \\
  && \exp\Bigl[\pm {\rm i} \bigl (q y_l-L\pi/2-\eta\ln(2qy_l)+\sigma_L\bigr ) \Bigr]\ ,
\end{eqnarray}
where $\sigma_L$ is the Coulomb phase-shift.
Therefore, $\Omega_{LS,JJ_z}^+$  ($\Omega_{LS,JJ_z}^-$) describe the
asymptotic outgoing (ingoing) $\pHe$ relative motion. Finally,
\begin{equation}
  \Psi_A^{LS,JJ_z}= \sum_{L^\prime S^\prime}
 \bigg[\delta_{L L^\prime} \delta_{S S^\prime} \Omega_{LS,JJ_z}^-
  - {\cal S}^{J\pi}_{LS,L^\prime S^\prime}
     \Omega_{L^\prime S^\prime,JJ_z }^+ \bigg] \ ,
  \label{eq:psia}
\end{equation}
where the parameters ${\cal S}^{J\pi}_{LS,L^\prime S^\prime}$ are the $S$-matrix
elements which determine phase-shifts and (for coupled channels) mixing parameters
at the energy $T_{c.m.}$.  Of
course, the sum over $L^\prime$ and $S^\prime$ is over  all values compatible
with the given $J$ and parity $\pi$. In particular, the sum over $L^\prime$
is limited to include either even or odd values such that $(-1)^{L^\prime}=(-)^L=\pi$.

The $S$-matrix elements ${\cal S}^{J\pi}_{LS,L^\prime S^\prime}$ and
coefficients $c^{LSJ}_\mu$ occurring in the HH expansion of $\Psi^{LS,JJ_z}_C$ are
determined by making the functional
\begin{equation}
   [{\cal S}^{J\pi}_{LS,L^\prime S^\prime}]=
    {\cal S}^{J\pi}_{LS,L^\prime S^\prime}
     -
        \left \langle\Psi^{L^\prime S^\prime,JJ_z}_{1+3} \left |
         H-E \right |
        \Psi^{LS,JJ_z}_{1+3}\right \rangle
\label{eq:kohn}
\end{equation}
stationary with respect to variations in the ${\cal S}^{J\pi}_{LS,L^\prime
S^\prime}$ and $c^{LSJ}_{\mu}$ (Kohn variational principle).
In the above equation, $E=T_{c.m.}-B(\He)$ is the energy of the system,
$B(\He)$ being the $\He$ binding energy.
By applying this principle, a linear set of equations
is obtained for ${\cal S}^{J\pi}_{LS,L^\prime S^\prime}$ and
$c^{LSJ}_{\mu}$. This linear system is solved using the Lanczos
algorithm.

This method can be applied in either coordinate or momentum
space, and using either local or non-local potentials~\cite{rep08}
(see also Ref.~\cite{Mea09} for an application to the $A=3$
scattering problem). The first step is a partial wave decomposition
of the asymptotic functions $\Omega_{LS,JJ_z}^\pm$, a task which can
be rather time consuming, in particular for the $J^\pi=2^-$
state. After this decomposition, the calculation of the matrix element
in Eq.~(\ref{eq:kohn}) is fast. Then, the problem reduces
to the solution of the linear system, which is performed using an
iterative method (however, this solution has to be repeated
several times due to the necessity to extrapolate the results, see
below).

The expansion of the scattering wave function in terms of the HH basis
is in principle infinite, therefore a truncation scheme is necessary.
The HH functions are essentially characterized by
the orbital angular momentum
quantum numbers $\ell_i$, $i=1,3$, associated with the three Jacobi
vectors, and the grand angular quantum number $K$ (each HH
function is a polynomial of degree $K$). The basis is
truncated to include states with $\ell_1+\ell_2+\ell_3\le \ell_{\rm
max}$ (with all possible re-coupling between angular and spin states appropriate to
the given $J$). Between these states, we retain only the HH functions with
$K\le K_{\rm max}$. In the calculation we have included
only states with total isospin $T=1$.

The numerical uncertainty comes from the numerical integrations needed
to compute the matrix elements of the Hamiltonian and the truncation
of the basis. It has been checked that the
numerical uncertainty of the calculated phase-shifts related to the numerical
integration is small (around $0.1$ \%). The NN interaction has been
limited to act on two-body states with total angular momentum $j\le
j_{\rm max}=8$ (at the considered energies, greater values of $j_{\rm
  max}$ are completely unnecessary).  The largest uncertainty is thus
related to the use of a finite basis. The convergence with
$\ell_{\rm max}$ is rather fast and the value $\ell_{\rm max}= 6$
have been found to be sufficient. The main problem is
related to the slow convergence of the results with
$K_{\rm max}$.  This problem can be partly overcome by performing
calculations for increasing values of $K_{\rm max}$ and then using some
extrapolation rule (see for example Ref.~\cite{Fisher06}) to get the
``$K_{\rm max}\rightarrow\infty$'' result. This procedure has an uncertainty which can be
estimated. A detailed study of this problem will be published
elsewhere~\cite{Vea11}. The convergence of the quantities of interest
in term of $K_{\rm max}$ is slower when NN potentials with a strong
repulsion at short interparticle distance are used such as for the AV18
potential. In this case we have estimated the uncertainty to be of the
order of $0.5$ \% in the extrapolated phase-shifts. This problem
is less relevant for the I-N3LO and the low-$k$ models. In these case,
the uncertainty has been estimated to be at most $0.3$ \%.

\subsection{FY Equations in Configuration Space}
\label{sec:FY}
In late sixties, Yakubovsky~\cite{Yaku_67} has managed to generalize the three-body equations
derived by Faddeev~\cite{Fad_60} to an arbitrary number of particles. These
equations were primarily derived for a system of particles submitted to short range
pair-wise potential $V^{s}$. Nevertheless it becomes possible to include also
repulsive Coulomb interaction if these, from now on called Faddeev-Yakubovsky
 equations, are formulated in configuration space.
To this aim, we split the Coulomb potential $V^{C}$ into two parts
(short and long range), $V^{C}=V^{s.C}+V^{l.C}$. The splitting procedure is quite
arbitrary, one should only take care that the long range part $V^{l.C}$ of the Coulomb
potential approaches sufficiently fast  the full Coulomb interaction $V^{C}$ when any of 
interparticle distances becomes large.
The simplest application of FY equations is the problem of four identical particles. They  result into a set of two differential equations
coupling the two so called FY components, namely $K_{12,3}^4$ and $H_{12}^{34}$, and have the form:
\begin{eqnarray}
   \lefteqn{ \left( E-H_{0}-V^s_{12}-\sum_{i<j} V^{l.C}_{ij}\right)
     K_{12,3}^4=\qquad\qquad} &&\nonumber\\
    &&\qquad(V^s_{12}+V^{s.C}_{12})P_1 \left[
     (1+\varepsilon P_{34})K_{12,3}^4+H_{12}^{34}\right]\ , \\
    \lefteqn{ \left( E-H_{0}-V^s_{12}-\sum_{i<j} V^{l.C}_{ij}\right)
      H_{12}^{34}=\qquad\qquad}&&\nonumber\\ 
     &&\qquad(V^s_{12}+V^{s.C}_{12})P_2 \left[
     (1+\varepsilon P_{34})K_{12,3}^4+H_{12}^{34}\right]\ ,  \label{FYE}
\end{eqnarray}%
where $P_1$, $P_2$ and $P_{34}$ are the particle permutation
operators,  equivalent to those described in Section~\ref{sec:AGS}, 
and  $\varepsilon=\pm1$ is a phase accounting for the Pauli principle
between two identical particles ($\varepsilon=+1$ for bosons and
$\varepsilon=-1$ for fermions).

%
Each FY component $F=(K,H)$ is  considered as a function of its proper set of
Jacobi~\cite{Rimas_03,Lazaus04} vectors $\bmx,\bmy,\bmz$ and expanded in angular
variables for each coordinate according to 
\begin{equation}\label{KPW}
  \langle\bmx,\bmy,\bmz|F\rangle=
  \sum_{\alpha} \; {F_{\alpha}(x,y,z)\over xyz} \;Y_{\alpha}
  (\hat{\bmx},\hat{\bmy},\hat{\bmz})\ .
\end{equation}
The quantities $F_{\alpha}$ are called regularized FY amplitudes and
$Y_{\alpha}$ are
tripolar harmonics, containing spin, isospin and angular momentum
variables.  The label $\alpha$ holds for the set of 10 intermediate
quantum numbers describing a $J^{\pi},T=1$ state in the partial wave basis.

The FY components $F=(K,H)$ are subject to Dirichlet-type boundary condition
imposed on a three dimensional rectangular grid.
Both components vanish on any of three $(x,y,z)$, axes,
as well as  at the borders $x=x_{max}$ and $y=y_{max}$ of the chosen resolution grid.
On contrary, a boundary condition equivalent to Eq. (\ref{eq:psia}) is imposed on the
 $z=z_{max}$ border for the FY components of type $K$; if no 2+2 particle
 channels are open FY component of type $H$ must also vanish at the $z=z_{max}$.

As discussed in the previous section, the expansion of the scattering wave function
in terms of the partial wave basis
is in principle infinite and a truncation scheme is necessary.
In this work the partial wave basis was
truncated to include all the states with $j_x\leq4$, $j_y\leq4$
and $j_z\leq3$, in the so-called $j-j$ coupling scheme~\cite{Rimas_03,Lazaus04}.
By studying the convergence of the calculated phase-shifts
with respect to the size of the partial wave basis, we have concluded that this
truncation scheme should provide results accurate at 1\% level.

The numerical implementation of these equations is described
in detail in Ref.~\cite{Rimas_03}.

\section{Results}
\label{sec:results}

In this section we present the phase-shifts for the most relevant
waves calculated using the three different methods described above.
The selected energies for $\nH$ are $E_n=1$, $2$, $3.5$ and $6$ MeV,
while for $\pHe$ are $E_p=2.25$, $4.05$ and $5.54$ MeV,
corresponding to cases where experiments have been carried out.

The states considered are those with
$J^\pi=0^\pm$, $1^\pm$, and $2^-$. The scattering in other
$J^\pi$ states is dominated by the centrifugal barrier and therefore the
phase-shifts are smaller and not very sensitive to the
interaction and the method used to calculate them.
Note that, for the $J^\pi=2^-$ state, we
have chosen to report only the ${}^3P_2$ phase-shift, since the
${}^3F_2$ phase-shift and the relative mixing parameter are in any
case very small. Nevertheless the coupling between the ${}^3P_2$ and ${}^3F_2$ waves
has been included in the
calculations, since the presence of the ${}^3F_2$ component
in the asymptotic part of the wave function has a
sizable effect on the ${}^3P_2$ phase-shift.

Let us remember that the $S$-matrix for elastic $\nH$ and $\pHe$
scattering has dimension 1 for $J^\pi=0^\pm$ states and
dimension $2$ for $J>0$. In the first case, the $S$-matrix is
parametrized as usual as ${\cal S}^{J\pi}_{LS,LS}=\exp(2i\delta^{J\pi}_{LS})$.
For $J>0$, since the $S$-matrix is
unitary and symmetric, we can write it as
\begin{eqnarray}
 S=O^T S_D O\ ,\label{eq:OSO}
\end{eqnarray}
with $S_D$ a diagonal matrix written as
\begin{eqnarray}
 (S_D)_{LS,L'S'}=\delta_{LL'}\delta_{SS'}
 e^{2i\delta^{J\pi}_{LS}}\ ,
\end{eqnarray}
where  $\delta_{LS}^{J\pi}$ is the phase-shift (in the
Blatt-Biederharn representation) of the wave $LS$. Due to
the unitarity properties,  $\delta_{LS}^{J\pi}$ is a real number. The matrix $O$ in
Eq. (\ref{eq:OSO}) is parametrized as
\begin{eqnarray}
 O & = & \begin{bmatrix}\cos\epsilon^{J\pi} & & \sin\epsilon^{J\pi}\\
  -\sin\epsilon^{J\pi} & & \cos\epsilon^{J\pi}
   \end{bmatrix} \ ,
\end{eqnarray}
where  $\epsilon^{J\pi}$ is the so called mixing parameter
of the given $J^\pi$ state. Clearly the values
of the phase-shifts and mixing parameters may depend on the
(arbitrary) choice on the coupling scheme between the spin of the two
clusters and the spherical harmonic function $Y_L({\bm y})$ in the
asymptotic functions $\Omega_{LS}^\pm$ (see, for example,
Eq.~(\ref{eq:psiom})). It can be shown that the phase-shifts defined 
as discussed above are independent on such choices, while the mixing
parameter, on the contrary, depends on them. Nevertheless, 
it is easy to establish the linear relation to transform the mixing
parameter from one coupling scheme to another. In the following, we
chose to report the mixing parameters defined in the $LS$ coupling
scheme by Eq.~(\ref{eq:psiom}). Moreover, in the following tables,
the values reported in a column labeled as ${}^{2S+1}L_J$ (using the
``spectroscopic notation'') are relative to the phase-shift
$\delta_{LS}^{J\pi}$.

In Table \ref{tab:nt-n3lo} we present the phase-shifts,
mixing parameters, and total cross sections
for $\nH$ scattering obtained using the I-N3LO potential
at the selected energies. By inspecting the table, we can notice the good
agreement between the three different techniques. 
The maximal deviation of the results is less than 1\%,
fully in line with the estimated errors. Furthermore, the agreement between
the results of AGS and HH techniques is even better, only in a
few cases the HH and AGS results differ by more than
0.5\%. The strongest deviation, of the order
of 0.4 deg, is observed with the FY results at the
largest studied energy. This slightly larger
deviation might be due to the necessity by the FY method to perform
the transformation of the aforementioned potential to configuration
space. In this respect, we note that the AGS calculation is fully
performed in momentum space, while in the HH calculation, part of the
needed matrix elements are calculated in momentum space (those
involving $\Psi^{LS,JJ_z}_C$, which have to be calculated with more
accuracy), and part in configuration space (those involving
$\Psi^{LS,JJ_z}_A$).

In Table~\ref{tab:nt-av18} we present the same $\nH$ results obtained
using the AV18 potential. In this case, the
convergence of the HH expansion is more problematic, in particular due
to the necessity to extrapolate the HH results. We observe that we
still have a very good agreement for the ${}^1S_0$, ${}^3P_0$,
${}^3P_2$, and $1^+$ phase-shifts, while the differences in the $1^-$
phase-shifts appear to be more enhanced. 

In Table~\ref{tab:nt-lowk} we have reported the phase-shifts obtained
with the low$-k$ potential derived from CD-Bonn. In this case, the
calculations has been performed using the AGS and HH methods, only. We
again observe an overall good agreement between the results obtained
by the two techniques, except for the lowest
energy where the differences are sizeable.

The total cross sections $\sigma_t$ are found to be in agreement within 0.05 b at
all considered energies. By comparing the values obtained using the
different potentials, we can observe the following well-known
characteristics: {\it (i)} at low energies the I-N3LO and AV18 models
overpredicts the experimental cross section. For example, at $E_n=1$ MeV,
$\sigma_t^{\rm expt}\approx 1.6$ b~\cite{PBS80}, while $\sigma_t^{\rm
  I-N3LO}\approx \sigma_t^{\rm AV18}\approx1.8$ b. On the contrary, with the
low-$k$ potential the calculated $\sigma_t$ is quite close to the
experimental one. This behavior is related to the strict relation
between the total cross section at low energy and the triton binding
energy~\cite{VKR98,Cie98,PHH01,Vea09}. {\it (ii)} at the peak
(around $E_n=3.5$ MeV), the experimental cross section has been
measured to be $\sigma_t^{\rm expt}\approx 2.45$ b~\cite{PBS80}. In
this case we note that the AV18 and low-$k$ potential models
underpredict sizeably the experimental value, while  $\sigma_t^{\rm
  I-N3LO}$ is quite close to it.

Let us now consider $\pHe$ scattering. The phase-shifts and mixing
parameters obtained within the three methods have been reported in
Tables~\ref{tab:ph-n3lo},~\ref{tab:ph-av18}, and ~\ref{tab:ph-vlowk},
corresponding respectively to the I-N3LO, AV18, and the low-$k$ NN potential
models. Here the differences between the various techniques
are larger than in the $\nH$ case, especially at low energy and for
the $J^\pi=0^\pm$ states. For the AV18 potential, we note that the HH
results are moreless intermediate between the AGS and FY results.

In Table~\ref{tab:ph-vlowk} we have also reported the phase-shifts and
mixing parameters obtained by the recent PSA~\cite{Dan10}. Note that
the low-$k$ potential used in this work is the only potential which
does not underestimate the $\Heq$ binding energy. The PSA estimates
have rather large errors. However, it is possible to draw some conclusions about
the capability of this low-$k$ potential model to describe the
experimental data.
As can be seen, the PSA S-wave phase-shifts seem to be well reproduced
(except for the ${}^1S_0$ phase-shift at $5.54$ MeV) by the
calculations. Also the ${}^3P_0$ and  ${}^1P_1$ agree well, but for
these cases the experimental errors are large. On the other hand, we
note a sizeable underestimation of the large ${}^3P_1$ and ${}^3P_2$
phase-shifts.

Let us now see how the fairly good agreement found for the
phase-shifts and mixing parameters calculated with the three different
methods reflects on the observables.
We have considered the differential cross
section and the neutron (proton) analyzing power $A_{y0}$ for $\nH$
($\pHe$) elastic scattering at the considered energies,
as functions of the c.m. scattering angle. Furthermore, we have also
considered the triton ($\He$) analyzing power $A_{0y}$. This
observable is in fact rather sensitive to
small variations of the phase-shifts
in the kinematical regime considered in this paper.

In Figs.~\ref{fig:n3lo-nt} and \ref{fig:n3lo-ph} we have reported the
results obtained using the AGS equation (solid lines),
the HH expansion method (dashed lines), and the FY equations (dotted
lines) using the I-N3LO potential. As can be seen by inspecting the
two figures, the three curves almost always perfectly coincide and
cannot be distinguished. We have also reported
the experimental data for the $\nH$ differential cross
section~\cite{SCS60} and the three $\pHe$
observables~\cite{Fam54,Mcdon64,Fisher06,All93,Vea01,Dan10}.
We note that the differences between the three calculations, where they
can be appreciated, are in any case always smaller than the
experimental errors.

The agreement between the three
calculations when the AV18 potential is adopted is again rather
satisfactory, as can be seen in Figs.~\ref{fig:av18-nt} and
\ref{fig:av18-ph}. A small disagreement can be observed only for the
$A_{0y}$ observable (see the panels in the last row of
Fig.\ref{fig:av18-ph}). This observable is also rather sensitive to
the small D-wave and F-wave phase-shifts not reported in
Tables~\ref{tab:nt-av18} and~\ref{tab:ph-av18}. 
We already know that the AV18 model contains a stronger repulsion at short
interparticle distance than the I-N3LO. As discussed above, the
convergence of the HH method for this case is more problematic and
consequently the calculations have a larger uncertainty. In spite of these
difficulties, the agreement in the considered observables is
still quite good.

Let us consider now the low-$k$ potential, which has no
repulsion at short interparticle distance.
Consequently, in this case, we expect a good agreement between
the results of the different techniques.  For this potential, the
calculations have been performed using the AGS (solid curves) and HH
(dashed curves) methods, only, and the
corresponding results are reported in Figs.~\ref{fig:vlowk-nt} and
\ref{fig:vlowk-ph}. The two curves
are practically indistinguishable, confirming that for soft potentials
the convergence of the calculations is excellent.

Finally, in the literature for $\pHe$ scattering, there exist
measurements of other spin correlation observables ($A_{yy}$,
$A_{xx}$, $A_{zz}$, $A_{xz}$, and $A_{zx}$). Also for these
observables we have found a good agreement between the
predictions obtained by the three different methods, for all the
potential models considered here. The comparison
of the theoretical predictions and the experimental data for these
observables will be discussed in the next section.

\section{Comparison with experimental data}
\label{sec:comp-theo-expt}

In this section we discuss the comparison between the theoretical
calculations and the experimental data. We consider here only $\pHe$
scattering since for this process the experimental data are more
abundant and precise. The figures presented in this section can be
considered as an update of previous
comparisons~\cite{Lea05,Fisher06,DF07,DF07b,Vea10}.
For the three observables considered so far
($d\sigma/d\Omega$, $A_{y0}$, and $A_{0y}$), the comparison between
theory and experiment can be inferred already from
Figs.~\ref{fig:n3lo-nt}--\ref{fig:vlowk-ph}. However, in order to
better appreciate the differences in the predictions obtained
by the three potential models as compared to the experimental data, we
summarize again in Fig.~\ref{fig:fig7} the results for
$d\sigma/d\Omega$, $A_{y0}$, and $A_{0y}$.
In order to take into account the (slight) different predictions
obtained using the three different theoretical methods,
we have decided to present the calculated observables for each
potential as bands. Each band contains 
the results obtained by using the three different methods.
As can be seen from Fig.~\ref{fig:fig7}, the differential unpolarized
cross sections obtained using the I-N3LO potential (red bands) agree
well with the experimental data. With the other two potentials we
observe some disagreement, in particular around $\theta_{\rm
c.m.}\approx 30$ deg and in the large scattering angle region. 
The results obtained for $A_{y0}$  are found to depend
on the potential model. Here, we observe the well known underprediction
of the experimental data by the theoretical
calculations. Interestingly, the results obtained with the low-$k$
potential are in a better agreement with the experimental $A_{y0}$. A
similar situation is found also for $A_{0y}$, as can be seen
in the three lower panels of Fig.~\ref{fig:fig7}. It is worthy
to note that the effect of supplementing the AV18 potential with the Urbana 3N force
model~\cite{Urb} has been found to be almost negligible for this
observable~\cite{Fisher06}. The inclusion of the new chiral 3N
potential derived in Ref.~\cite{N07} is under study~\cite{Vea11} (see
Ref.~\cite{Vea09} for a preliminary report).

In Fig.~\ref{fig:fig8}, we report the results
found for the $A_{yy}$ and $A_{xx}$ spin correlations
at the three different proton energies. As can be seen, for
these two observables the predictions obtained with the three
potentials are almost identical. We observe that the calculated
$A_{yy}$ is slightly at
variance with respect to the experimental data, while the $A_{xx}$
observable is reasonably well reproduced by the calculations.

Finally, in Fig.~\ref{fig:fig9}  we compare the results
obtained for the $A_{xz}$, $A_{zx}$, and $A_{zz}$ spin correlation
observables. In this case, only the 
$E_p=5.54$ proton laboratory energy is considered, since only for this energy
experimental data exist. Also in this case, the sensitivity to
the different potential models is not significant. Moreover, the calculations
reproduce well the (few) experimental data.

\section{Conclusions}
\label{sec:conc}

In this work, we have studied several low energy $\nH$ and $\pHe$
elastic observables by using three different approaches, the HH, AGS and
FY techniques. Around four years ago, some of the authors of the
present paper presented very accurate solutions of the 4-nucleon
scattering problem using the AGS
technique~\cite{DF07,DF07b,DF07c}. They were able to take
into account the long-range Coulomb interaction using the
screening-renormalization method~\cite{Alt78,DFS05}.  In recent years,
also the accuracy of the calculations performed using the HH and FY
techniques increased~\cite{Vea09,Vea10,Lazaus09}. Therefore, it
becomes appropriate to compare the results obtained by the
different methods in order to
test their capability to solve the 4N scattering problem. This is the
primary aim of the present paper. Another
important motivation is to
provide a set of solid converged results in the literature,
which could represent useful benchmarks for future applications in
$A=4$ scattering.

In the present paper we have shown that for I-N3LO and the
selected low-$k$ potential model, which have a ``soft'' repulsion at short interparticle
distances (the low-$k$ model has no repulsion at all), the
results obtained by the different techniques are in very good
agreement. With the AV18 potential, the agreement is not so
perfect, although the (slight) differences can be appreciated only for some
small polarization observables. We can conclude therefore, that the
$A=4$ scattering problem is nowadays solved with a very good
accuracy, better than 1\%.

Concerning the comparison with the experimental data, we have
confirmed the large underprediction of the $\pHe$ $A_{y0}$ observable,
a problem already put in evidence some time ago~\cite{Cie98,Fon99,Vea01},
and certainly related to the $N-d$ ``$A_y$ puzzle''. For this
observable we have observed a moderate dependence on the
considered potential models.
The low-$k$ potential is found to give a better description of the observable when
compared with the experimental data. However, the same potential does
not reproduce well the unpolarized cross section. We have also found a small
underprediction of the theoretical $A_{0y}$ and
$A_{yy}$ observables, while other measured observables, such as
$A_{xx}$, $A_{xz}$, $A_{zx}$, and $A_{zz}$ spin correlation
coefficients, show less sensitivity to the potential models. They are
in good agreement with the available (sparse) experimental data.

The discrepancies found, in particular for $A_{y0}$,  indicate a
serious difficulty of the existing NN force models in describing the
4N continuum. This difficulty can hardly be solved by the inclusion of a
standard type 3NF, used to reproduce the few-nucleon
binding energies~\cite{Lea05,Fisher06,Vea10}. Its origin could
rather lie either in the NN
forces themselves, or in the presence of a 3NF of unknown type.
Clearly, an eventual solution of the $A=4$ $A_{y0}$ problem should be
related in some way to the solution of the $N-d$ ``$A_y$
puzzle''.

Finally, we conclude noticing that it
would be interesting to extend the present analysis to $\pH$, $\nHe$
and $\dd$ scattering observables, which have already been calculated in the
framework of the AGS equations for different NN
interactions~ \cite{DF07c,DF10}.

\bigskip
{\bf Acknowledgment:}

The FY work  was performed using the HPC resources of IDRIS under the
allocation 2010-i2009056006 
made by GENCI (Grand Equipement National de Calcul Intensif). We thank
the staff members of the IDRIS for their constant help.

\bibliographystyle{prsty}

\begin{table*}[t]
\begin{ruledtabular}
\begin{tabular}{*{13}{r}}
$E_n$ & ${}^1S_0$  & ${}^3P_0$ & ${}^3S_1$ & ${}^3D_1$ &  $\epsilon^{1+}$ &
  ${}^1P_1$  & ${}^3P_1$ & $\epsilon^{1-}$ &
   ${}^3P_2$  & $\sigma_t $
\\  \hline
1.0 & -38.10 &  4.15 & -33.32 & -0.09 & -0.23 & 5.99  &  9.63 &  9.44 &  8.98 & 1.77 & AGS \\
    & -38.02 &  4.10 & -33.31 & -0.08 & -0.22 & 5.86  &  9.64 &  9.14 &  8.95 & 1.77 & HH\\
    & -38.31 &  4.00 & -33.56 & -0.11 & -0.24 & 6.13  & 10.13 &  9.6  &  9.16 & 1.81 & FY\\
\hline
2.0 & -51.93 & 10.54 & -45.66 & -0.36 & -0.44 & 13.13 & 24.18 &  9.15 & 23.96 & 2.13 & AGS \\
    & -51.98 & 10.50 & -45.72 & -0.35 & -0.43 & 13.12 & 24.25 &  9.18 & 23.96 & 2.13 & HH\\
    & -52.34 & 10.54 & -45.99 & -0.39 & -0.50 & 13.55 & 25.15 &  9.62 & 24.52 & 2.19 & FY\\
\hline
3.5 & -65.54 & 20.31 & -57.99 & -0.91 & -0.72 & 20.74 & 40.94 &  9.45 & 43.98 & 2.38 & AGS \\
    & -65.66 & 20.26 & -58.08 & -0.91 & -0.72 & 20.94 & 40.97 &  9.55 & 43.91 & 2.38 & HH \\
    & -66.15 & 20.62 & -58.40 & -0.91 & -0.79 & 21.17 & 41.50 &  9.33 & 44.42 & 2.41 & FY \\
\hline
6.0 & -80.53 & 32.71 & -71.75 & -1.77 & -1.16 & 26.88 & 52.35 & 10.62  & 60.04 & 1.97 & AGS \\
    & -80.57 & 32.55 & -71.79 & -1.80 & -1.15 & 26.92 & 52.25 & 10.68  & 60.01 & 1.97 & HH\\
    & -80.98 & 33.40 & -71.93 & -1.81 & -1.22 & 27.05 & 52.00 & 10.71  & 59.96 & 1.97 & FY\\
\end{tabular}
\end{ruledtabular}
\caption{ \label{tab:nt-n3lo}
$\nH$ phase-shifts and mixing parameters (in degrees) and total cross section
$\sigma_t$ (in barns) for  the I-N3LO potential  at  $E_n = 1.0$, 2.0, 3.5, and 6.0 MeV.}
\end{table*}

\begin{table*}[t]
\begin{ruledtabular}
\begin{tabular}{*{13}{r}}
$E_n$ & ${}^1S_0$  & ${}^3P_0$ & ${}^3S_1$ & ${}^3D_1$ &  $\epsilon^{1+}$ &
  ${}^1P_1$  & ${}^3P_1$ & $\epsilon^{1-}$ &
   ${}^3P_2$  & $\sigma_t $
\\  \hline
1.0 & -38.52 &  4.36 & -33.67 & -0.10 & -0.24 &  6.15 &  9.64 &  9.38 & 8.94 & 1.80 & AGS \\
    & -38.44 &  4.26 & -33.57 & -0.09 & -0.21 &  5.87 &  9.44 &  9.19 & 8.82 & 1.78 & HH \\
    & -38.55 &  4.36 & -33.75 & -0.09 & -0.28 &  6.14 &  9.62 &  9.45 & 8.93 & 1.81 & FY \\
\hline
2.0 & -52.43 & 10.93 & -46.08 & -0.38 & -0.46 & 13.30 & 23.90 &  8.99 & 23.45 & 2.12 & AGS \\
    & -52.41 & 10.82 & -46.04 & -0.37 & -0.42 & 13.00 & 23.39 &  9.19 & 23.21 & 2.10 & HH \\
    & -52.55 & 10.92 & -46.23 & -0.37 & -0.47 & 13.36 & 23.86 &  9.07 & 23.44 & 2.13 & FY\\
\hline
3.5 & -66.12 & 20.75 & -58.48 & -0.93 & -0.75 & 20.73 & 40.09 &  9.24 & 42.51 & 2.33 & AGS \\
    & -66.14 & 20.61 & -58.53 & -0.95 & -0.72 & 20.68 & 39.63 &  9.48 & 42.22 & 2.32 & HH\\
    & -66.23 & 20.62 & -58.66 & -0.94 & -0.77 & 20.75 & 39.98 &  9.31 & 42.37 & 2.33 & FY\\
\hline
6.0 & -81.03 & 32.77 & -72.19 & -1.78 & -1.22 & 26.53 & 51.13 & 10.37 & 57.87 & 1.93 & AGS \\
    & -81.05 & 32.61 & -72.40 & -1.87 & -1.20 & 26.55 & 51.27 & 10.57 & 57.94 & 1.93 & HH \\
    & -80.95 & 32.53 & -72.22 & -1.86 & -1.24 & 26.58 & 50.95 & 10.47 & 57.57 & 1.92 & FY \\
\end{tabular}
\end{ruledtabular}
\caption{ \label{tab:nt-av18}
Same as Table~\ref{tab:nt-n3lo}, but for the AV18 potential.}
\end{table*}

\begin{table*}[b]
\begin{ruledtabular}
\begin{tabular}{*{13}{r}}
$E_n$ & ${}^1S_0$  & ${}^3P_0$ & ${}^3S_1$ & ${}^3D_1$ &  $\epsilon^{1+}$ &
  ${}^1P_1$  & ${}^3P_1$ & $\epsilon^{1-}$ &
   ${}^3P_2$  & $\sigma_t $
\\  \hline
1.0 & -36.39 &  3.52 & -32.03 & -0.08 & -0.19 &  5.34 &  8.86 &  9.79 &  8.37 & 1.62 & AGS \\
    & -36.08 &  3.41 & -31.88 & -0.06 & -0.19 &  5.01 &  8.70 &  9.34 &  8.23 & 1.60 & HH \\
\hline
2.0 & -49.73 &  9.03 & -43.99 & -0.32 & -0.37 & 12.05 & 22.61 &  9.76 &  22.79 & 1.96 & AGS \\
    & -49.61 &  8.94 & -43.95 & -0.28 & -0.37 & 11.83 & 22.51 &  9.69 &  22.71 & 1.95 & HH \\
\hline
3.5 & -62.94 & 17.75 & -56.01 & -0.82 & -0.63 & 19.72 & 39.30 & 10.24 &  43.20 & 2.26 & AGS \\
    & -63.06 & 17.74 & -56.10 & -0.79 & -0.63 & 19.88 & 39.41 & 10.32 &  43.25 & 2.27 & HH \\
\hline
6.0 & -77.57 & 29.44 & -69.51 & -1.66 & -1.03 & 26.38 & 51.44 & 11.57 &  60.41 & 1.94 & AGS \\
    & -77.77 & 29.46 & -69.64 & -1.70 & -1.04 & 26.56 & 51.48 & 11.63 &  60.45 & 1.94 & HH \\
\end{tabular}
\end{ruledtabular}
\caption{ \label{tab:nt-lowk}
Same as Table~\ref{tab:nt-n3lo}, but  for the
low-momentum potential derived from the CD Bonn potential. In this
case, only the AGS and HH results are reported.}
\end{table*}

\begin{table*}[b]
\begin{ruledtabular}
\begin{tabular}{*{13}{r}}
$E_p$ & ${}^1S_0$  & ${}^3P_0$ & ${}^3S_1$ & ${}^3D_1$ &  $\epsilon^{1+}$ &
  ${}^1P_1$  & ${}^3P_1$ & $\epsilon^{1-}$ &   ${}^3P_2$
\\  \hline
2.25 & -40.64 &  8.04 & -35.00 & -0.24 & -0.53 & 10.64 & 17.29 & 8.61 & 16.26  & AGS \\
     & -41.23 &  7.73 & -35.47 & -0.34 & -0.54 & 10.42 & 17.11 & 8.69 & 16.11  & HH \\
     & -41.57 &  7.74 & -35.49 & -0.28 & -0.58 & 10.84 & 17.75 & 8.43 & 16.41  & FY \\
\hline
4.05 & -58.23 & 17.94 & -50.79 & -0.94 & -0.84 & 18.90 & 35.50 & 8.73 & 36.61 & AGS \\
     & -58.61 & 17.76 & -51.01 & -0.97 & -0.82 & 18.97 & 35.43 & 8.85 & 36.53 & HH \\
     & -59.12 & 18.12 & -51.15 & -0.96 & -0.94 & 19.26 & 35.78 & 8.62 & 36.88 & FY \\
\hline
5.54 & -68.28 & 25.41 & -60.02 & -1.45 & -1.08 & 23.05 & 44.54 & 9.28 & 48.53 & AGS \\
     & -68.50 & 25.07 & -60.11 & -1.51 & -1.07 & 23.00 & 44.34 & 9.36 & 48.29 & HH \\
     & -69.00 & 25.81 & -60.03 & -1.40 & -1.18 & 23.16 & 44.13 & 9.28 & 48.33 & FY \\
\end{tabular}
\end{ruledtabular}
\caption{ \label{tab:ph-n3lo}
$\pHe$ phase-shifts and mixing parameters (in degrees)
for the I-N3LO potential  at  $E_p = 2.25$, 4.05, and 5.54 MeV.}
\end{table*}

\begin{table*}[t]
\begin{ruledtabular}
\begin{tabular}{*{13}{r}}
$E_p$ & ${}^1S_0$  & ${}^3P_0$ & ${}^3S_1$ & ${}^3D_1$ &  $\epsilon^{1+}$ &
  ${}^1P_1$  & ${}^3P_1$ & $\epsilon^{1-}$ &
   ${}^3P_2$
\\  \hline
2.25 & -41.11 &  8.46 & -35.26 & -0.26 & -0.56 & 10.93 & 17.35 & 8.43 & 16.29  & AGS \\
     & -41.53 &  7.84 & -35.65 & -0.34 & -0.48 & 10.33 & 16.76 & 8.56 & 15.76  & HH \\
     & -41.70 &  7.82 & -36.01 & -0.29 & -0.52 & 10.52 & 17.08 & 9.01 & 15.90 & FY \\
\hline
4.05 & -58.70 & 18.41 & -51.17 & -0.93 & -0.87 & 18.89 & 34.83 & 8.46 & 35.65 & AGS \\
     & -58.93 & 17.89 & -51.34 & -0.98 & -0.82 & 18.85 & 33.49 & 8.79 & 35.33 & HH \\
     & -59.02 & 17.61 & -51.76 & -1.00 & -0.82 & 18.57 & 34.81 & 8.82 & 35.36 & FY \\
\hline
5.54 & -68.75 & 25.82 & -60.41 & -1.43 & -1.12 & 22.91 & 43.65 & 9.00 & 47.09 & AGS \\
     & -68.96 & 25.05 & -60.78 & -1.55 & -1.10 & 22.89 & 43.20 & 9.32 & 46.72 & HH\\
     & -68.92 & 24.74 & -60.91 & -1.55 & -1.06 & 21.93 & 44.01 & 8.08 & 46.53 & FY \\
\end{tabular}
\end{ruledtabular}
\caption{ \label{tab:ph-av18}
Same as Table~\ref{tab:ph-n3lo}, but  for the AV18 potential.}
\end{table*}

\begin{table*}[b]
\begin{ruledtabular}
\begin{tabular}{*{13}{r}}
$E_p$ & ${}^1S_0$  & ${}^3P_0$ & ${}^3S_1$ & ${}^3D_1$ &  $\epsilon^{1+}$ &
  ${}^1P_1$  & ${}^3P_1$ & $\epsilon^{1-}$ &   ${}^3P_2$
\\  \hline
2.25 & -38.74 &  6.85 & -33.60 & -0.21 & -0.46 &  9.63 & 15.95 &  9.13 & 15.13  & AGS \\
     & -38.96 &  6.53 & -33.80 & -0.26 & -0.43 &  9.21 & 15.76 &  8.98 & 14.91 & HH \\
     & -39.1$\pm$1.7 & 5 $\pm$ 6 & -34.5$\pm$0.7 & & & 8$\pm$2 &
     17$\pm$4 &10$\pm$20 & 16.5$\pm$0.7 & PSA \\
\hline
4.05 & -55.74 & 15.60 & -48.93 & -0.87 & -0.75 & 17.79 & 34.64 & 9.43 & 35.19 & AGS \\
     & -56.09 & 15.48 & -49.15 & -0.85 & -0.74 & 18.02 & 33.93 & 9.54 & 35.33 & HH \\
     & -56.3$\pm$0.6 & 14.1$\pm$0.9 & -49.3$\pm$0.5 & & & 17.3$\pm$1.6 &
     34.9$\pm$0.3 &13$\pm$2 & 37.6$\pm$0.6 & PSA \\
\hline
5.54 & -65.54 & 22.54 & -57.97 & -1.36 & -0.98 & 22.30 & 43.13 & 10.08 & 47.78  & AGS \\
     & -65.97 & 22.57 & -58.18 & -1.39 & -0.97 & 22.49 & 43.21 & 10.13 & 47.73 & HH \\
     & -67.8$\pm$0.9 & 21.3$\pm$0.7 & -58.6$\pm$0.3 & & & 21.2$\pm$1.7 &
     45.2$\pm$0.5 &14$\pm$2 & 51.5$\pm$0.5 & PSA \\
\end{tabular}
\end{ruledtabular}
\caption{ \label{tab:ph-vlowk}
Same as Table~\ref{tab:ph-n3lo}, but for the low-$k$ potential.
In this Table, only the AGS and HH results are reported. The phase-shifts and mixing
parameters obtained by the recent PSA~\protect\cite{Dan10} are also shown.}
\end{table*}

\begin{widetext}

\newpage

\begin{figure}[htb]
\begin{center}
\includegraphics[scale=0.7]{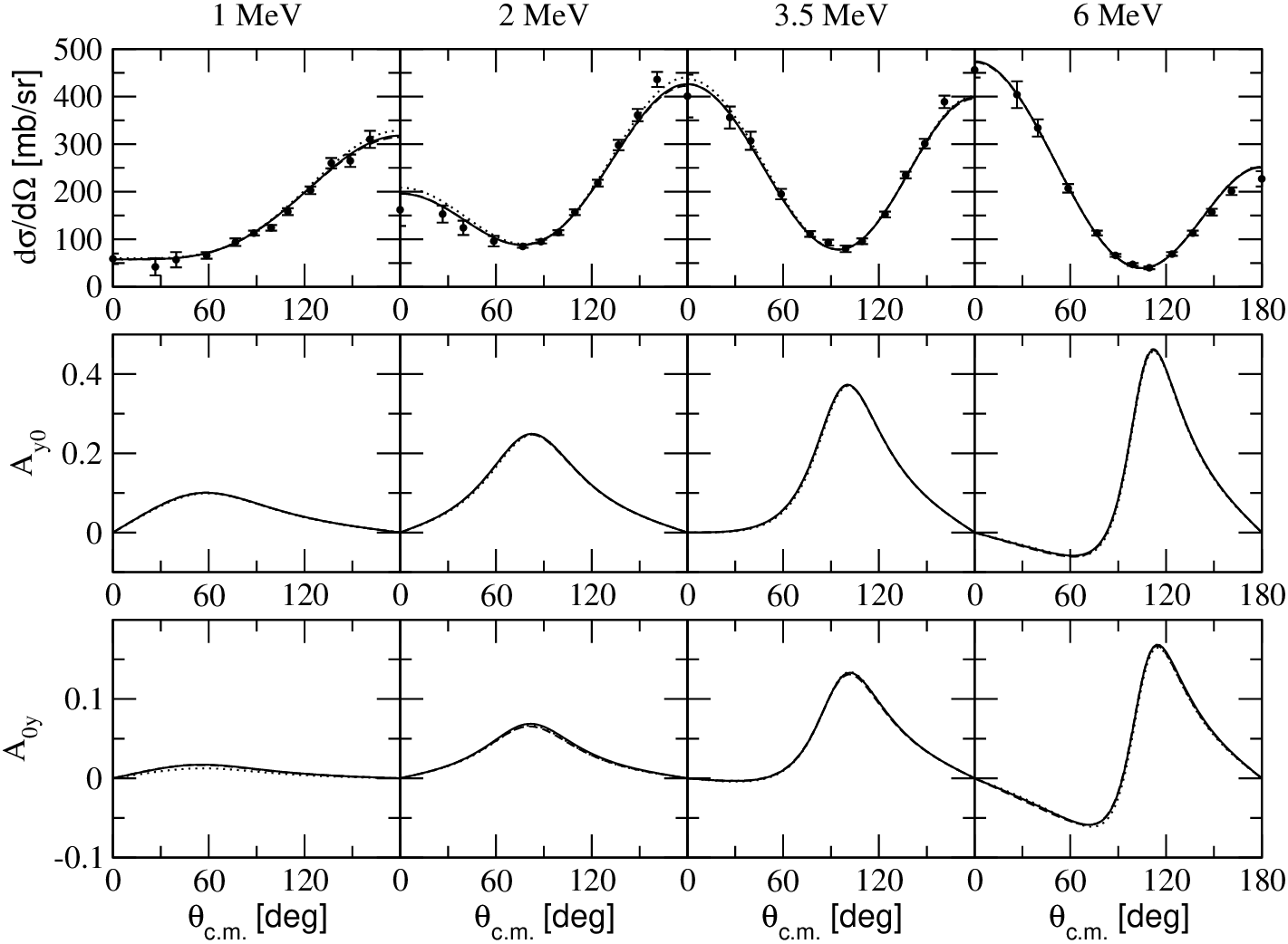}
\end{center} 
\caption{ \label{fig:n3lo-nt} Differential cross
section and neutron and triton analyzing powers $A_{y0}$ and
$A_{0y}$ for $\nH$ elastic scattering at $E_n=1$, $2$, $3.5$, and $6$ MeV
neutron lab energies as functions of the c.m. scattering
angle. Results obtained using the AGS equation (solid lines),
the HH expansion method (dashed lines), and the FY equations (dotted
lines) using the I-N3LO potential are compared. For most of the cases the three curves coincide
and cannot be distinguished.
The experimental data are from Ref.~\cite{SCS60}.}
\end{figure}

\begin{figure}[htb]
\begin{center}
\includegraphics[scale=0.7,clip]{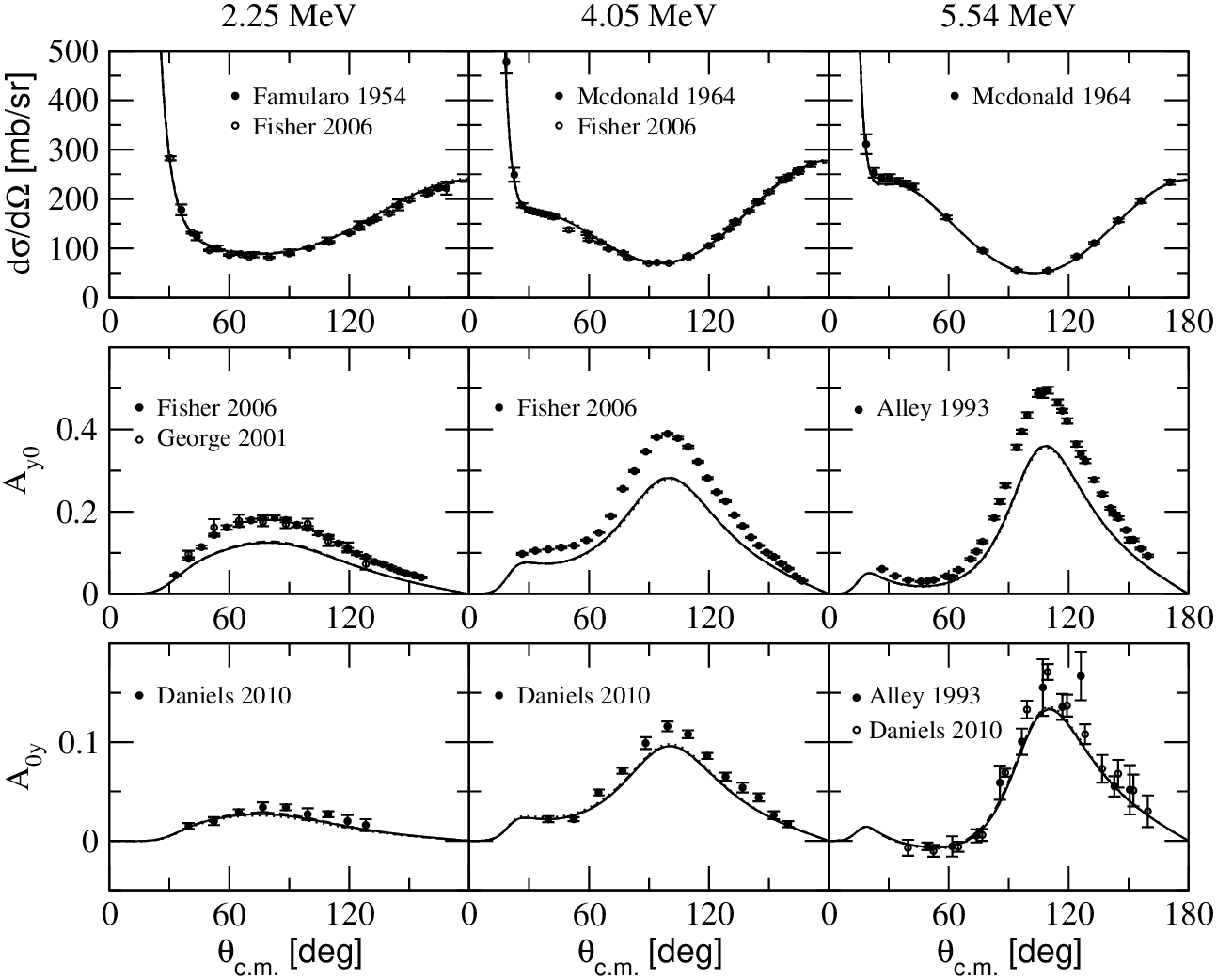}
\end{center} 
\caption{ \label{fig:n3lo-ph} Same as Fig.~\protect\ref{fig:n3lo-nt},
but for $\pHe$ elastic scattering at $E_p=2.25$, $4.05$, and $5.54$ MeV
proton lab energies. The experimental data are from
Refs.~\cite{Fam54,Mcdon64,Fisher06,All93,Vea01,Dan10}.}
\end{figure}

\begin{figure}[htb]
\begin{center}
\includegraphics[scale=0.7,clip]{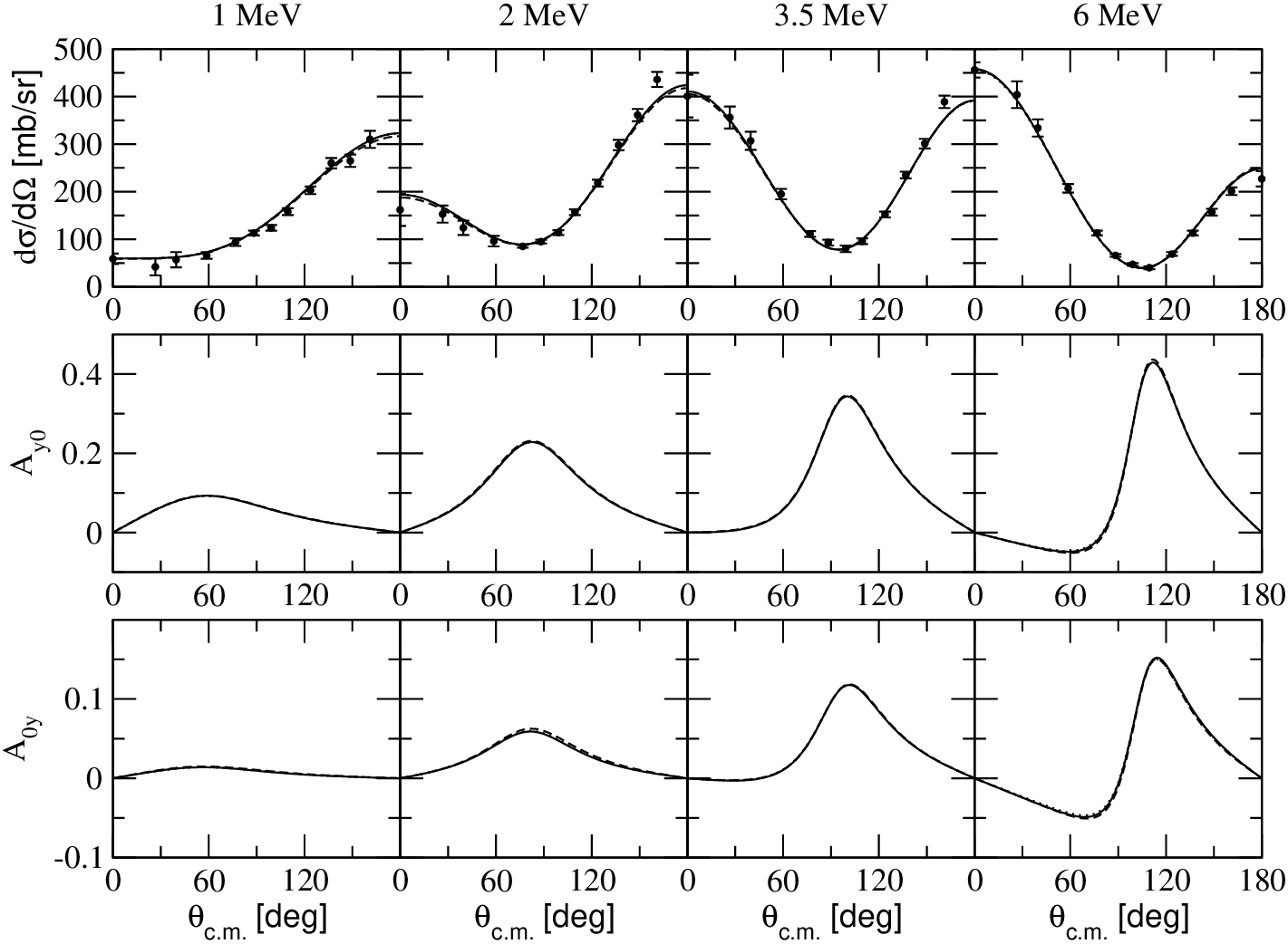}
\end{center} 
\caption{ \label{fig:av18-nt} Same as Fig.~\protect\ref{fig:n3lo-nt},
  but for the AV18 potential.}
\end{figure}

\begin{figure}[htb]
\begin{center}
\includegraphics[scale=0.7,clip]{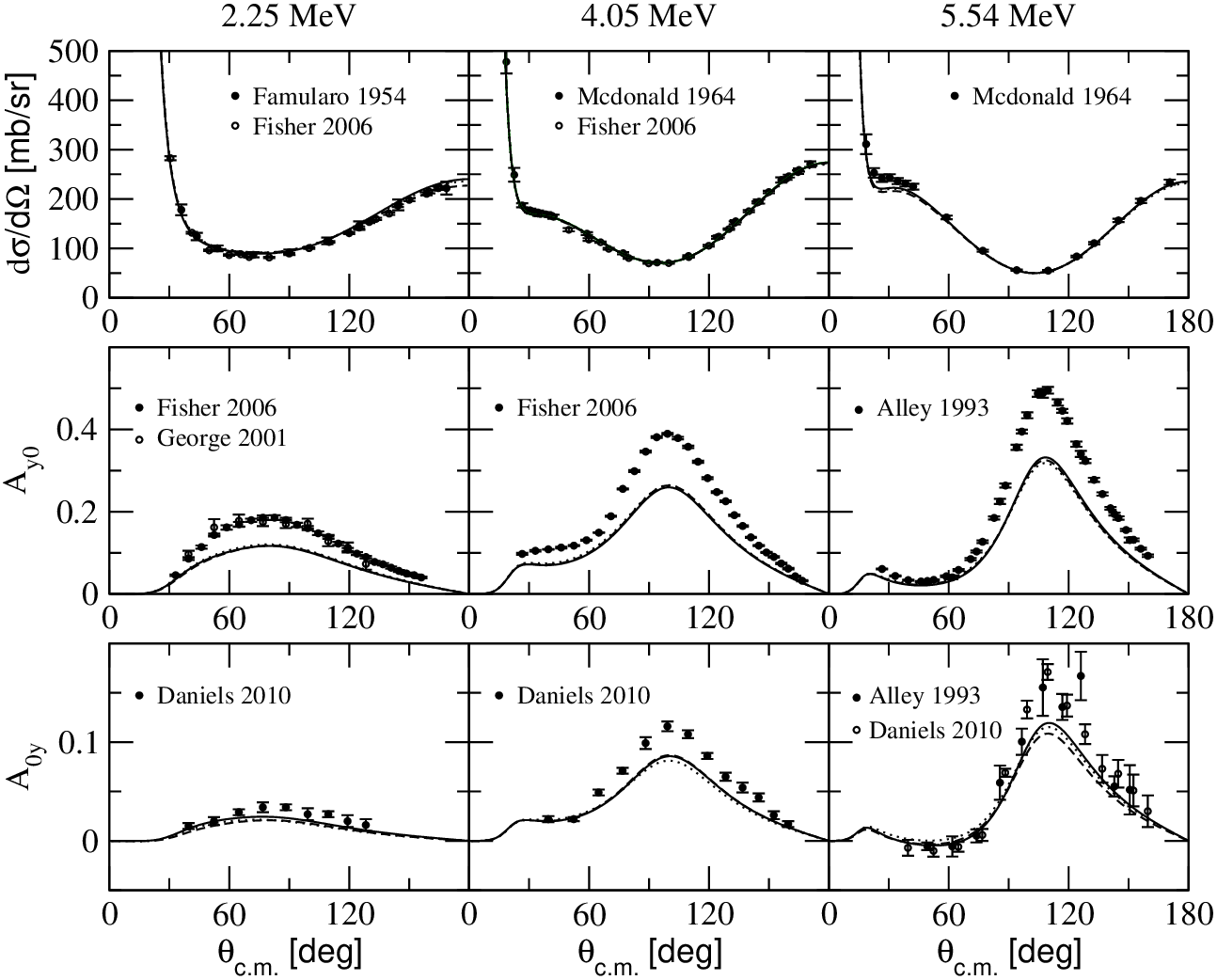}
\end{center} 
\caption{ \label{fig:av18-ph} Same as Fig.~\protect\ref{fig:n3lo-ph},
  but for the AV18 potential.}
\end{figure}

\begin{figure}[htb]
\begin{center}
\includegraphics[scale=0.7,clip]{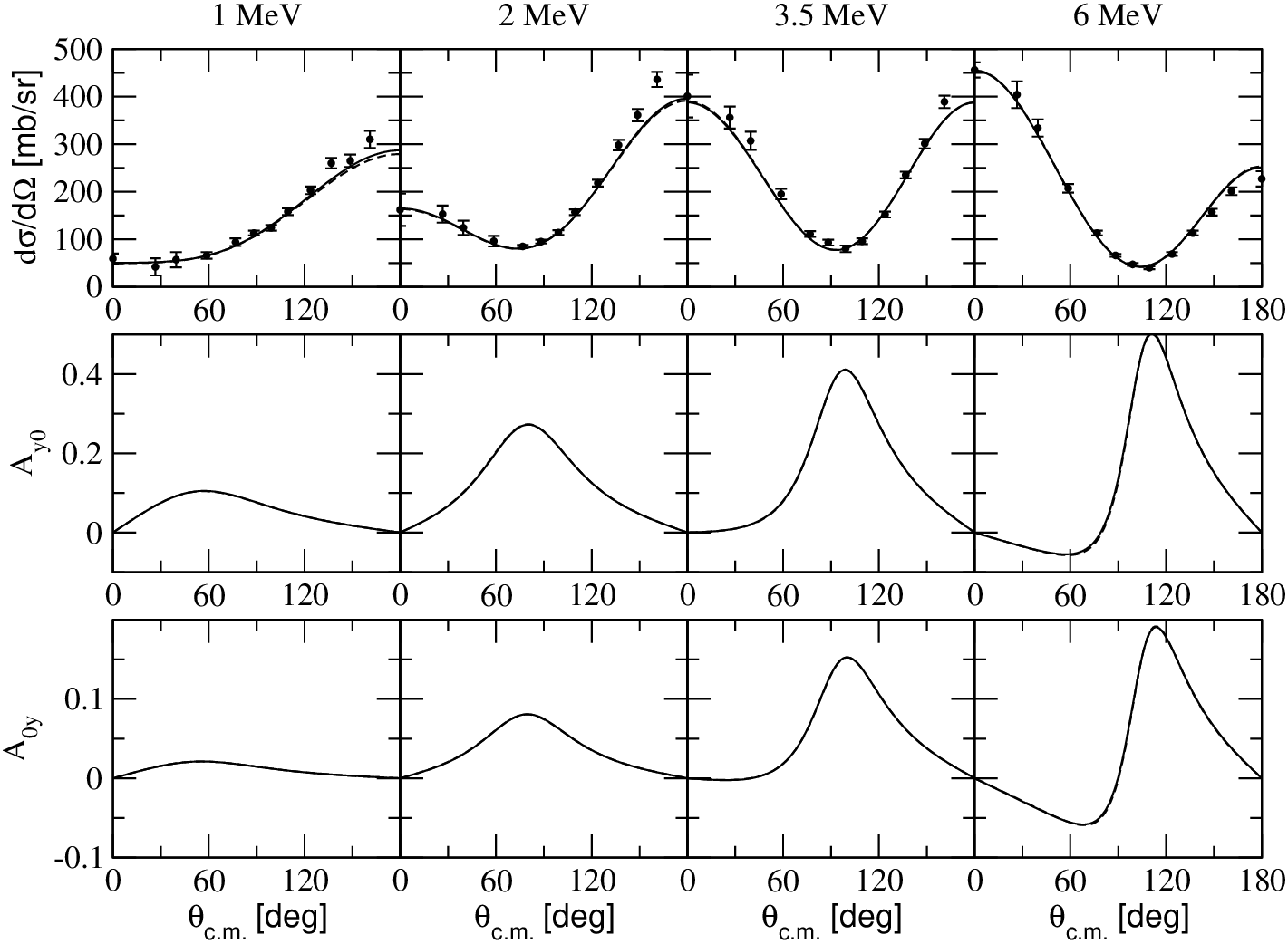}
\end{center} 
\caption{ \label{fig:vlowk-nt} Same as Fig.~\protect\ref{fig:n3lo-nt},
  but for the low-$k$ potential. Only the AGS and HH results are reported.}
\end{figure}

\begin{figure}[htb]
\begin{center}
\includegraphics[scale=0.7,clip]{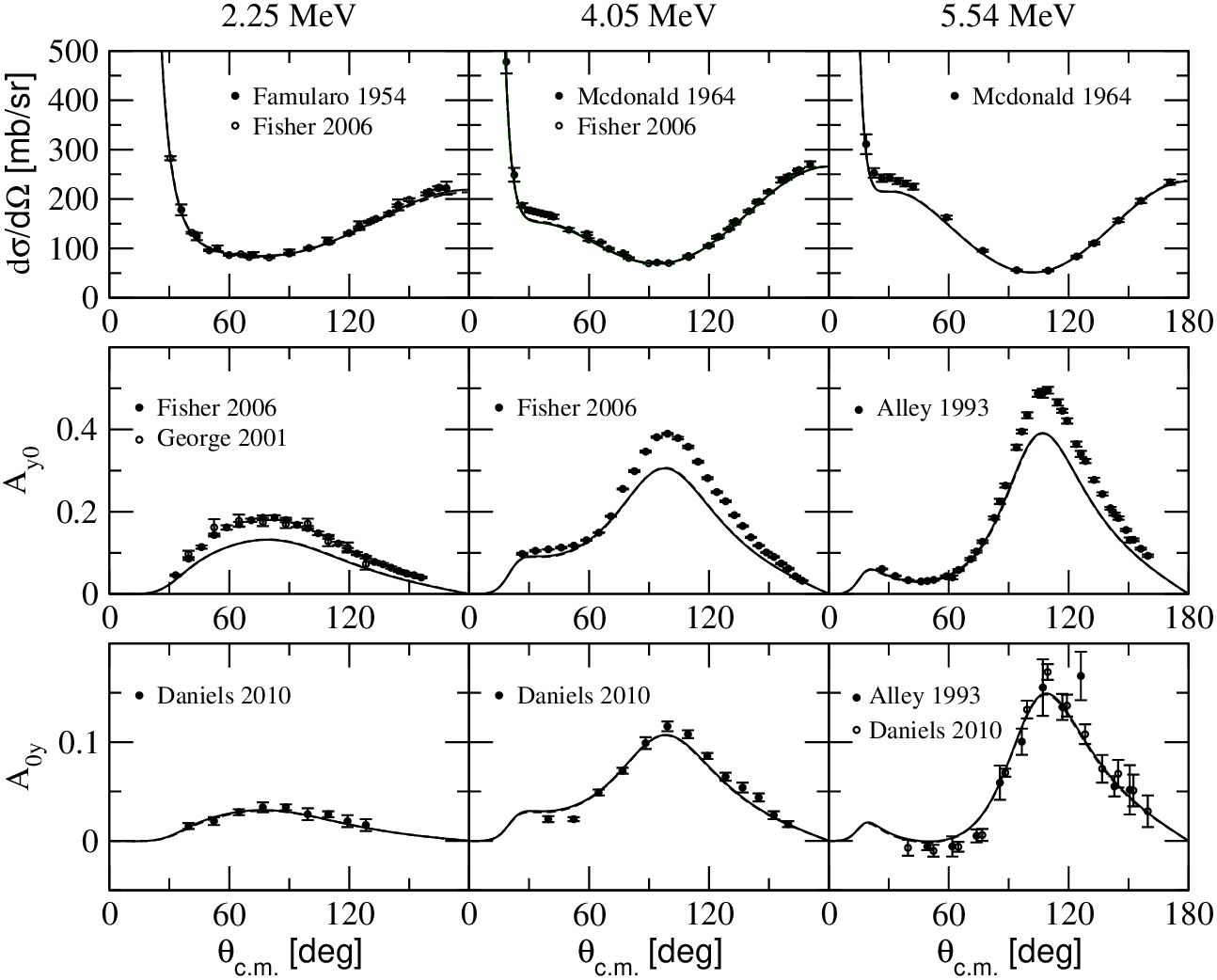}
\end{center} 
\caption{ \label{fig:vlowk-ph} Same as Fig.~\protect\ref{fig:n3lo-ph},
  but for the low-$k$ potential. Only the AGS and HH results are reported.}
\end{figure}

\begin{figure}[htb]
\begin{center}
\includegraphics[scale=0.6,clip]{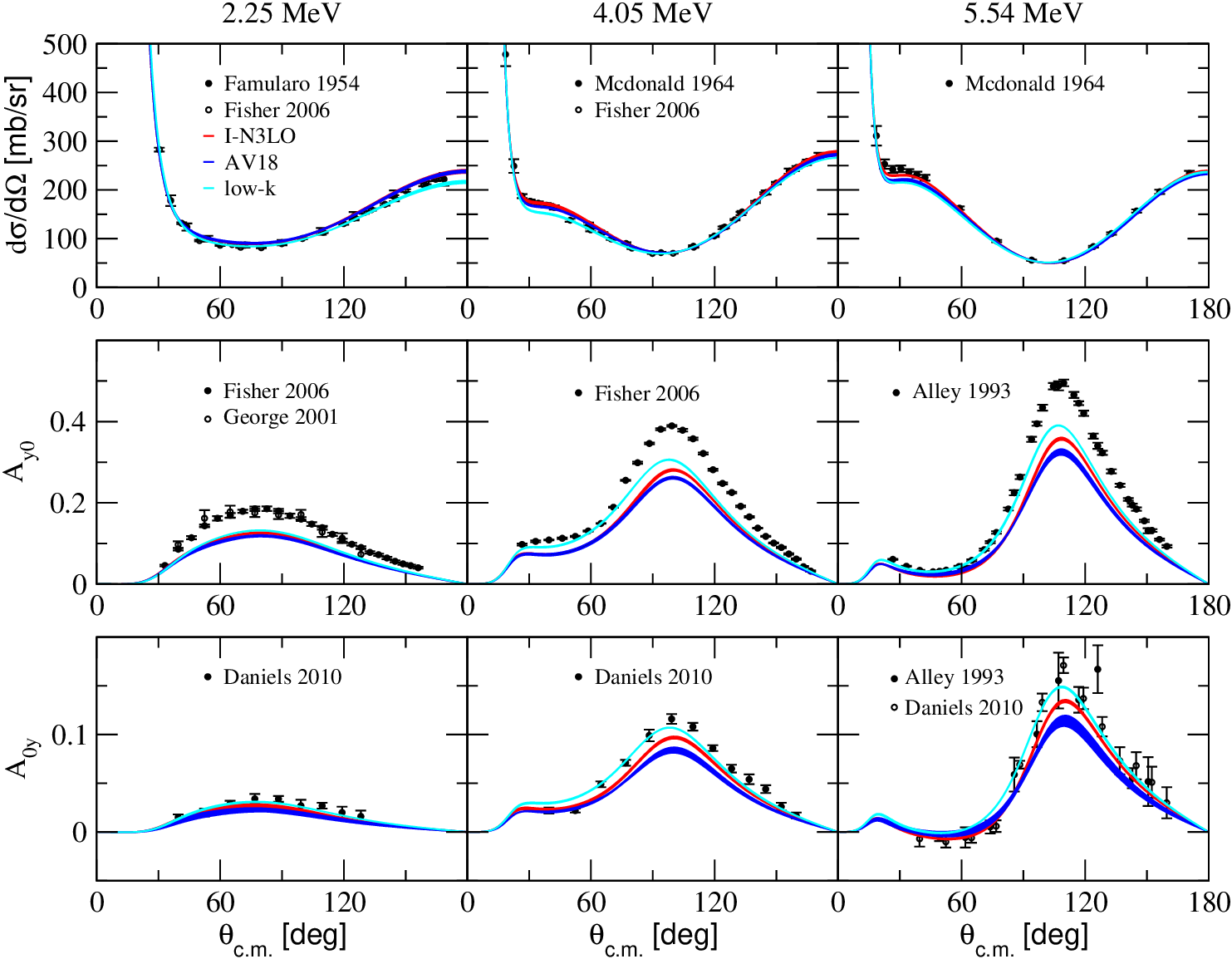}
\end{center} 
\caption{ \label{fig:fig7} (Color online) Differential cross
section, proton analyzing power, and $\He$ analyzing power for $\pHe$
elastic scattering at  $E_p=2.25$, $4.05$, and $5.54$ MeV 
proton lab energies obtained using the I-N3LO (red bands),
AV18 (blue bands), and the low-$k$ (cyan bands) potential models. The
experimental data are from 
Refs.~\cite{Fam54,Mcdon64,Fisher06,All93,Vea01,Dan10}.}
\end{figure}

\begin{figure}[htb]
\begin{center}
\includegraphics[scale=0.6,clip]{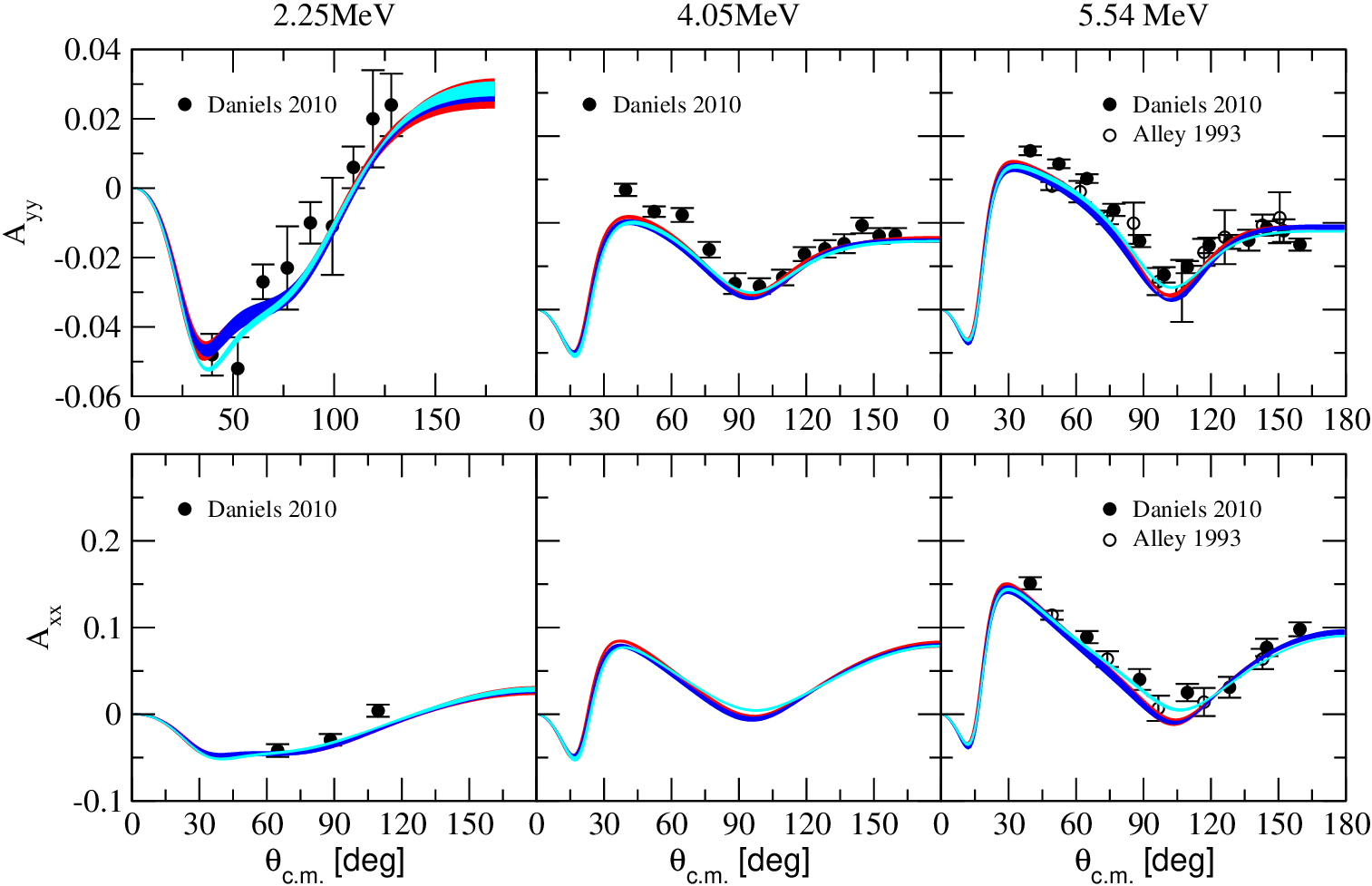}
\end{center} 
\caption{ \label{fig:fig8} (Color online) Same as Fig.~\protect\ref{fig:fig7}, but for
the spin correlation $A_{yy}$ and $A_{xx}$ observables. The
experimental data are from Refs.~\cite{All93,Dan10}.}
\end{figure}

\begin{figure}[htb]
\begin{center}
\includegraphics[scale=0.6,clip]{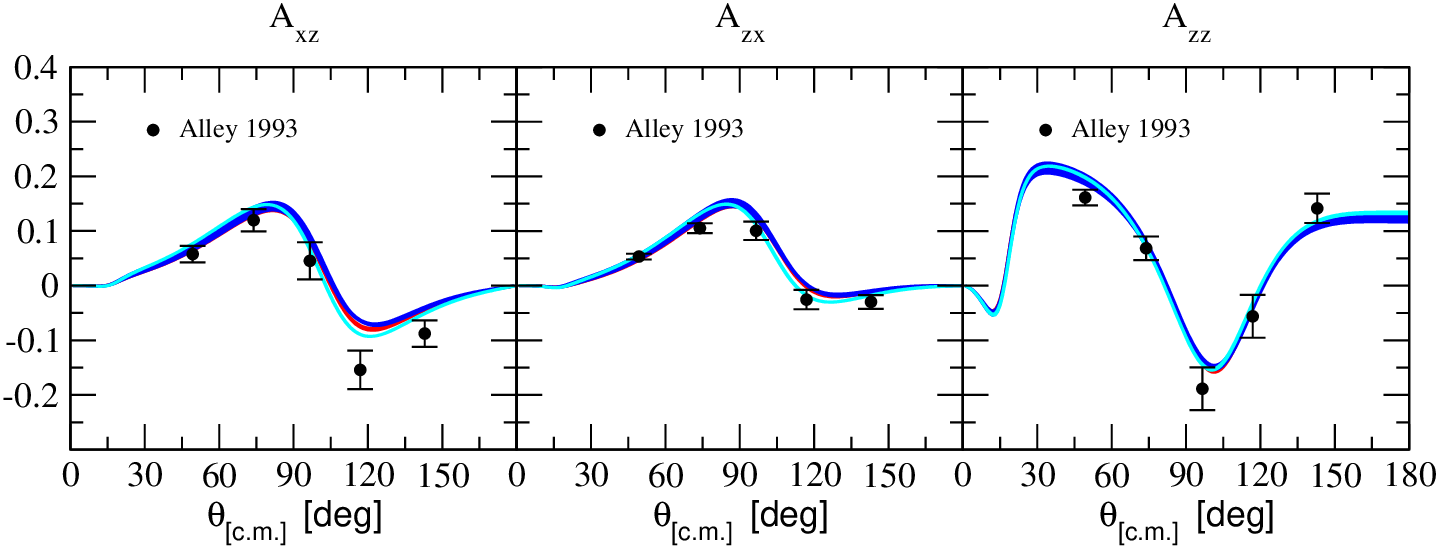}
\end{center} 
\caption{ \label{fig:fig9} (Color online) Same as Fig.~\protect\ref{fig:fig7}, but for
the spin correlation $A_{xz}$, $A_{zx}$, and $A_{zz}$ observables
(at $E_p=5.54$ MeV proton lab energy, only). The experimental data are from
Ref.~\cite{All93}.}
\end{figure}
\end{widetext}

\end{document}